\documentclass[aps,prl,twocolumn,amsmath,amssymb,superscriptaddress,10pt,nobibnotes]{revtex4-2}
\pdfoutput=1

\usepackage{amsfonts, amssymb, amsmath,mathrsfs}
\usepackage{graphicx}
\usepackage{color}
\usepackage[utf8]{inputenc}
\usepackage[T1]{fontenc}
\usepackage[svgnames]{xcolor}
\usepackage{textcomp}

\usepackage{hyperref}
\hypersetup{colorlinks,urlcolor=DarkBlue,linkcolor=DarkBlue,citecolor=DarkBlue,pdfdisplaydoctitle=true,pdfpagemode=UseOutlines,bookmarksnumbered=true,bookmarksopen=true} 

\makeatletter
\renewcommand{\eqref}[1]{\textup{\tagform@{\ref*{#1}}}}
\makeatother

\renewcommand{\bm}[1]{\boldsymbol{\mathbf{#1}}}

\newcommand{\im}{\operatorname{Im}}

\newcommand{\eq}[1]{Eq.~\eqref{#1}}

\newcommand{\Eq}[1]{Equation~\eqref{#1}}

\newcommand{\fig}[1]{Fig.~\ref*{#1}}

\newcommand{\beginsupplement}{%
        \setcounter{table}{0}
        \renewcommand{\thetable}{S\arabic{table}}%
        \setcounter{figure}{0}
        \renewcommand{\thefigure}{S\arabic{figure}}%
        \setcounter{section}{0}
        \renewcommand{\thesection}{\Roman{section}}
        \setcounter{equation}{0}
        \renewcommand{\theequation}{S\arabic{equation}}
     }

\begin{document}

\title{Influence of the Local Scattering Environment on the\\Localization Precision of Single Particles}

\author{Dorian Bouchet}
\altaffiliation{\href{mailto:d.f.bouchet@uu.nl}{d.f.bouchet@uu.nl}}
\affiliation{Nanophotonics, Debye Institute for Nanomaterials Science, Utrecht University, P.O. Box 80000, 3508 TA Utrecht, the Netherlands}
\author{Rémi Carminati}
\affiliation{Institut Langevin, ESPCI Paris, PSL University, CNRS, 1 rue Jussieu, 75005 Paris, France}
\author{Allard P. Mosk}
\affiliation{Nanophotonics, Debye Institute for Nanomaterials Science, Utrecht University, P.O. Box 80000, 3508 TA Utrecht, the Netherlands}

\begin{abstract}
We study the fundamental limit on the localization precision for a subwavelength scatterer embedded in a strongly scattering environment, using the external degrees of freedom provided by wavefront shaping. For a weakly scattering target, the localization precision improves with the value of the local density of states at the target position. For a strongly scattering target, the localization precision depends on the dressed polarizability that includes the back action of the environment. This numerical study provides new insights for the control of the information content of scattered light by wavefront shaping, with potential applications in sensing, imaging, and nanoscale engineering.
\end{abstract}

\maketitle

Localizing a small object in a complex scattering environment using wave scattering is a widespread problem in many fields, including material and life sciences. For instance, in nanofabrication, it is essential to control the manufacturing of structured samples and notably to localize defects in micro-electro-mechanical systems (MEMS)~\cite{osten_optical_2018}, semiconductor chips~\cite{vogel_technology_2007,alexanderliddle_lithography_2011} or photonic crystals~\cite{koenderink_optical_2005}. In life sciences, studying the inner structure of the cell implies the localization of nanoparticles or fluorophores in scattering environments, for instance in particle tracking experiments~\cite{zijlstra_single_2011,taylor_interferometric_2019}. Multiple scattering of acoustic waves or microwaves also complicates indoor localization of emitting or scattering devices~\cite{deak_survey_2012,xiao_survey_2016}. Yet, for many applications, characterizing complex scattering materials by solving the inverse problem is still possible thanks to the large amount of prior information available to the observer through design considerations~\cite{szameit_sparsity-based_2012,zhang_far-field_2016}. For this class of problems, defining and maximizing the information content of the data on a specific scattering object is a critical step in order to reach the best possible precision for imaging and metrology applications. 

Estimation theory provides a definition of the precision in the estimation of a parameter (for example the position of a target) through the Cramér-Rao inequality~\cite{kay_fundamentals_1993}. This inequality sets a lower bound to the variance of the estimated value of the parameter, known as the Cramér-Rao lower bound (CRLB). This bound depends on different features of the physical model, including the statistics of the measurement noise, the intrinsic properties of the scattering medium as well as the illumination and detection scheme. This theoretical limit has found useful applications in the design of optical imaging setups, for instance in the context of dynamic single-molecule measurements~\cite{watkins_information_2004}, diffuse optical imaging~\cite{boffety_analysis_2008} or lifetime measurements~\cite{kollner_how_1992,bouchet_cramer-rao_2019}. The CRLB has also been proposed to define the resolution of an imaging system~\cite{ram_beyond_2006,sentenac_influence_2007}. Furthermore, the concept is widely used to assess the localization precision in super-resolution imaging techniques based on single-molecule detection~\cite{ober_localization_2004,mortensen_optimized_2010,deschout_precisely_2014}. Recently, the idea arose that the localization precision of single molecules could be improved by spatially modulating either the incident or the emitted field to minimize the CRLB~\cite{shechtman_optimal_2014,balzarotti_nanometer_2017}. In parallel, advanced wavefront protocols were developed to control wave propagation in strongly scattering media~\cite{mosk_controlling_2012}, notably enabling the focusing of light waves inside materials~\cite{vellekoop_demixing_2008,hsieh_digital_2010,choi_measurement_2013,ma_time-reversed_2014,ambichl_focusing_2017,ruan_focusing_2017}. It is plausible that the localization precision for a hidden target can be improved by focusing light upon it, however this situation has not been rigorously analyzed so far.

In this Letter, we address this question by studying the fundamental limit in the precision on the localization of a subwavelength scatterer enclosed in a strongly scattering medium. We find out that the local environment of the target strongly influences the resulting localization precision. For a weakly scattering target (that is, when recurrent scattering between the target and the environment can be neglected), the key parameter driving the localization precision is the local density of states (LDOS), which is a fundamental quantity affecting many aspects of light-matter interaction such as spontaneous emission and thermal emission~\cite{barnes_fluorescence_1998,carminati_electromagnetic_2015}. For a strongly scattering target, the localization precision depends on the dressed polarizability of the scatterer, which describes the back action of the environment beyond the weak-coupling regime~\cite{bouchet_quantum_2019}. These results offer new insights to improve the performances of imaging and metrology techniques using wavefront shaping.

\begin{figure}[ht]
\centering
\includegraphics[width=7.2cm]{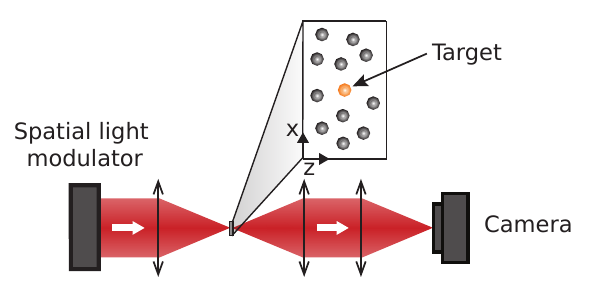}
\caption{Representation of a slab composed of several dipole scatterers and illuminated by a SLM. In all simulations, the thickness of the system is set to $L_z=10 \, \lambda $, and the transverse dimension is set to $L_x=50 \, \lambda$.}
\label{geometry}
\end{figure}

We consider a model system composed of two-dimensional scatterers arranged in a slab geometry, as represented in \fig{geometry}. One scatterer, located in the center of the system, is chosen as the target to be localized. The other scatterers, with random positions, define a complex scattering medium. In this way, we can investigate universal properties of the localization precision of the target, without being influenced by features specific to a given scattering nanostructure. This model of a scattering medium has been used for the description of basic problems in mesoscopic physics~\cite{de_vries_point_1998,rotter_light_2017}, up to the regime of Anderson localization~\cite{caze_strong_2013,skipetrov_absence_2014}. It is similar to that used in Ref.~\cite{irishina_source_2012} to study the inverse reconstruction of the position of fluorophores. In order to constrain the problem, we assume that only the position $\bm{r}_0=(x_0,z_0)$ of the target is unknown. The goal is then to estimate the two coordinates of the target using coherent illumination at a wavelength $\lambda=2 \pi/k$, where $k$ is the wavenumber in vacuum. We further assume that the incident field is either a plane wave or a sum of plane waves with equal amplitude and different incidence angles, as generated in practice by a phase-only spatial light modulator (SLM). The response of the subwavelength scatterers is described by an electric polarizability $\alpha$ and a scattering cross-section $\sigma_s=k^3 |\alpha|^2/4$. We denote the polarizability of the target by $\alpha_0$, and take its scattering cross-section to be $\lambda/1000$, ensuring that this scatterer is weakly coupled to its environment. We take the polarizability of the other scatterers at resonance ($\alpha=4i/k^2$), which is not an essential feature of the model but allows us to maximize their scattering cross-sections, and therefore to minimize the number of scatterers needed to reach the multiple-scattering regime. In order to compute the scattered field, we use the coupled dipole method, which is an exact formulation of the scattering problem in the limit of scatterers much smaller than the wavelength~\cite{lax_multiple_1952}. This method allows us to calculate the average (or expected) pixel intensity as measured by a camera located in the image plane of an ideal imaging system, which images the output plane $z=L_z$ (see Supplementary Section~I). 

Any measurement process is intrinsically probabilistic due to noise fluctuations that limit the precision on the determination of the position of the target in otherwise perfect conditions. Thus, the measured data must be described by a random variable $\bm{X}$. The joint probability density function $p(\bm{X} ; \bm{\theta})$ of the data set, parameterized by the set of unknown parameters $\bm{\theta}$ to be estimated, is used to define the Fisher information matrix~\cite{kay_fundamentals_1993}
\begin{equation}
\left[ \bm{\mathcal{F}}(\bm{\theta})\right]_{jk} = \left \langle \left[ \frac{\partial \ln p(\bm{X} ; \bm{\theta})}{\partial \theta_j} \right] \left[ \frac{\partial \ln p(\bm{X} ; \bm{\theta})}{\partial \theta_k} \right] \right \rangle \; .
\end{equation}
Here $\langle \, \dots \rangle $ denotes the average over noise fluctuations. While any noise statistics can be included in the formalism, we assume here that values measured on different pixels of the camera are statistically independent and follow a Poisson distribution, which corresponds to an experiment limited only by photon noise. The information matrix is then expressed by 
\begin{equation}
\left[ \bm{\mathcal{F}}(\bm{\theta})\right]_{jk} = \sum_{i=1}^{N} \frac{1}{I_i} \left( \frac{\partial I_i}{\partial \theta_j} \right) \left( \frac{\partial I_i}{\partial \theta_k} \right) \; ,
\label{eq_information_poisson}
\end{equation}
where $N$ is the total number of pixels and $I_i$ is the average value of the intensity measured by the $i$-th pixel. From \eq{eq_information_poisson}, we can compute the CRLB, which bounds the error in the determination of the parameter $\theta_j$, by
\begin{equation}
\mathcal{C}_{j}=\sqrt{\left[ \bm{\mathcal{F}}^{-1}(\bm{\theta})\right]_{jj}} \; .
\end{equation}
While there exists no general methodology to build an efficient estimation algorithm that reaches the CRLB, maximum likelihood estimation is the most popular approach to obtain practical estimators that are asymptotically efficient \cite{kay_fundamentals_1993}. Moreover, it is possible to obtain an explicit expression of such an estimator, in the limit of small parameter variations and for a large number of detected photons (see Supplementary Section~II). This estimator may be used to estimate the position of the target even when the positions of the
scatterers that constitute the scattering environment are known with some uncertainties (see Supplementary Section~III).

\begin{figure}[ht]
\centering
\includegraphics[width=7.4cm]{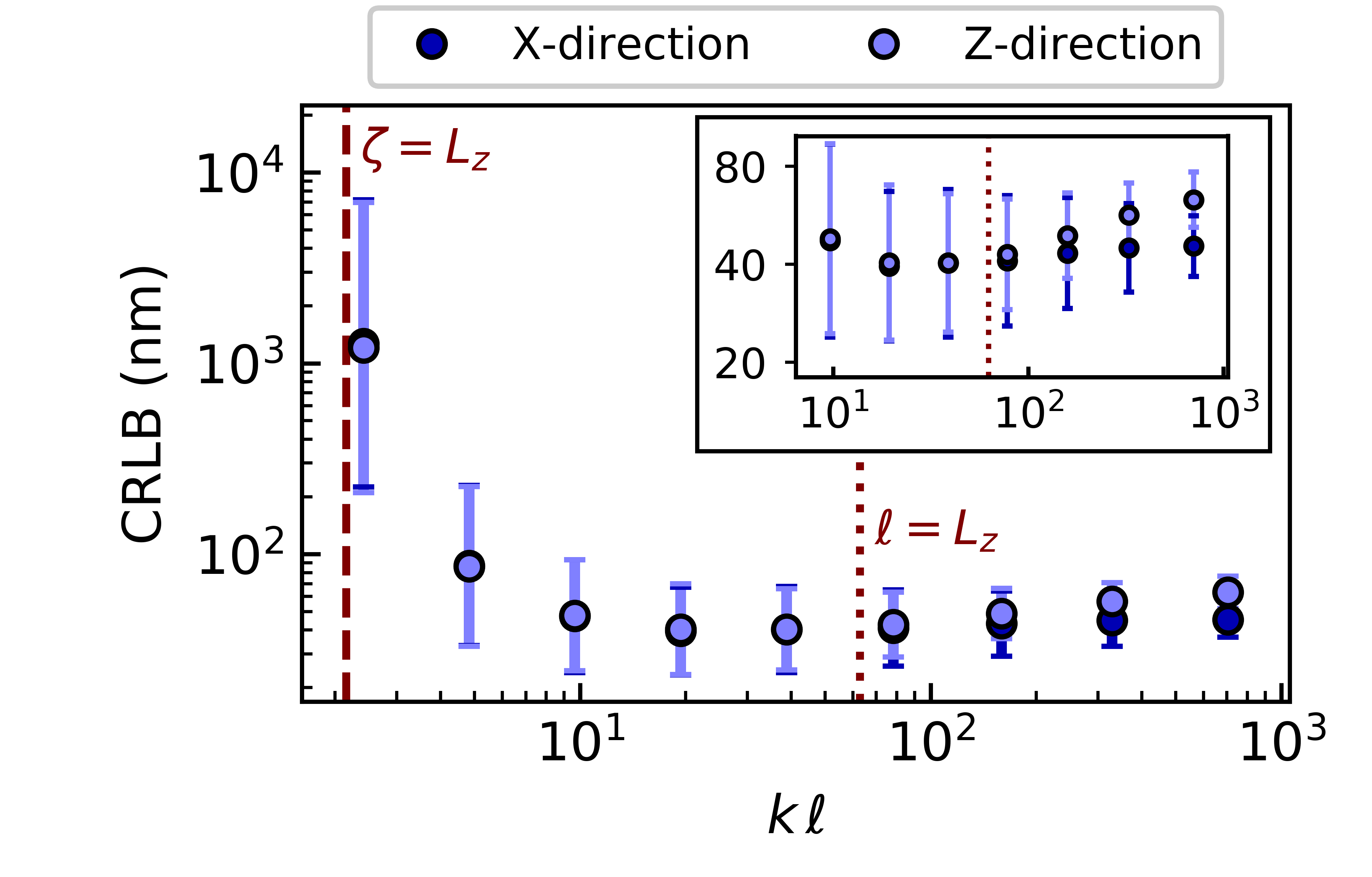}
\caption{CRLB as a function of the normalized scattering mean free path $k \ell$ for plane-wave illumination. Dotted lines correspond to $\ell=L_z$ (defining the transition to the multiple-scattering regime) and dashed lines correspond to $\zeta=L_z$ (defining the onset of Anderson localization). Each point represents the geometric mean over $1000$ configurations of the disordered medium, and error bars represent 1-sigma intervals. The inset shows the same data on a smaller scale.}
\label{fig_scaling}
\end{figure}

The CRLB can be evaluated in our model system by computing the average value of the intensity reaching the camera pixels using the coupled dipole method, and by evaluating the derivatives in \eq{eq_information_poisson} using a finite difference scheme. As only the coordinates of the target need to be estimated, we define $\bm{\mathcal{C}}=(\mathcal{C}_x, \mathcal{C}_z )$ where $\mathcal{C}_x$ and $\mathcal{C}_z$ are the CRLB on each coordinate. For the calculations, we choose $\lambda=633$~nm and an average incident intensity (integrated over the invariant $y$ coordinate) of $I_0=10^4$~photons per \textmu m. One can then easily deduce the CRLB for other values of $\lambda$ and $I_0$ by noting that the CRLB scales with $\lambda$ and with $I_0^{-1/2}$. In order to study the influence of multiple scattering on the precision in the estimation of the target position, we generate different random configurations of the medium that we illuminate with a plane wave at normal incidence, and we study the statistical distribution of the Cramér-Rao bound, with the statistics now performed with respect to disorder. Changing the density of scatterers $\rho_s$ allows us to modify the independent scattering (or Boltzmann) mean free path $\ell=(\rho_s \sigma_s)^{-1}$~\cite{lagendijk_resonant_1996}. In \fig{fig_scaling} we show the first two moments of the CRLB distribution as a function of $k \ell$. In the single-scattering regime ($\ell \gtrsim L_z$), the average CRLB depends on the coordinate to be estimated ($x_0$ or $z_0$), as expected for one isolated scatterer. In contrast, for $\ell<L_z$, the average CRLB is the same for both coordinates due to multiple scattering that restores isotropy. In this regime, the probability distribution of the CRLB follows a log-normal distribution (see Supplementary Section~IV), whose moments strongly depend on the scattering mean free path. We also observe that the average CRLB shows a minimum in this regime, demonstrating that on average multiple scattering improves the localization precision. Finally, when the localization length becomes on the order of the size of the medium, the CRLB strongly increases due to the onset of Anderson localization which suppresses light transmission~\cite{vollhardt_diagrammatic_1980} (we use $\zeta = \ell \exp (\pi k \ell/2)$ as a rough approximation of the localization length~\cite{gupta_localization_2003}).

\begin{figure}[ht]
\centering
\includegraphics[width=8.5cm]{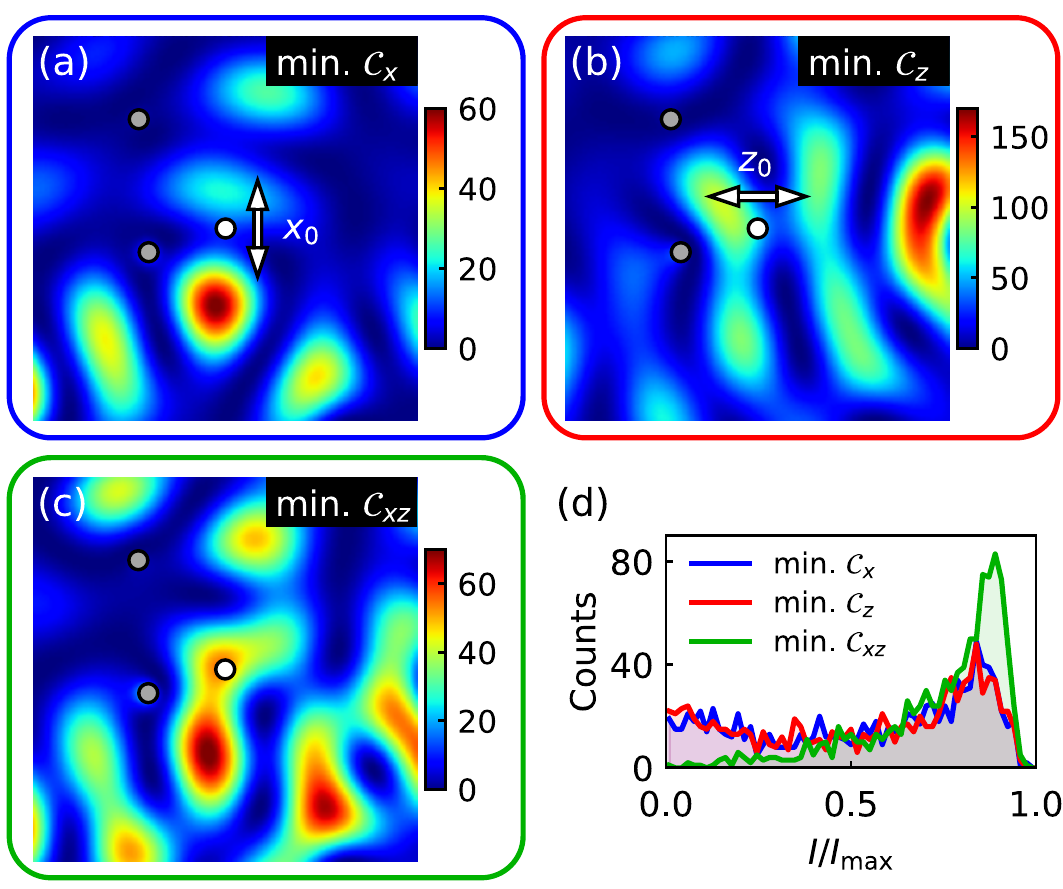}
\caption{Intensity $I/I_0$ around the target for a scattering medium illuminated by a wavefront optimized for (a)~the transverse coordinate $x_0$, (b)~the longitudinal coordinate $z_0$ and (c)~both coordinates simultaneously. An area of $2 \lambda \times 2 \lambda$ is displayed on each map. (d)~Histogram of the intensity ratio $I/I_{\mathrm{max}}$ at the target position when optimizing for $x_0$, $z_0$ and both coordinates simultaneously. }
\label{fig_intensity}
\end{figure}

The CRLB provides a figure of merit that can be optimized using the external degrees of freedom provided by wavefront shaping. In order to test the optimization of information in the presence of multiple scattering, we generate 1000 configurations of the medium in the diffusive regime ($k \ell = 9.7$, optical thickness $L_z/\ell=6.5$), assumed to be illuminated using a phase-only SLM composed of $N_e = 64$ elements. We then minimize the CRLB using a global optimization algorithm based on simulated annealing~\cite{kirkpatrick_optimization_1983}. The optimized field distribution weakly depends on the initial guess fed to the optimization algorithm (see Supplementary Section~V), which suggests that the obtained solutions are close to the global optimum. We show in \fig{fig_intensity}~(a) and~(b) the intensity around the target for a scattering medium illuminated by incident fields independently optimized for the determination of $x_0$ and $z_0$, respectively. The incident wavefront associated with the highest information content depends on the coordinate to be estimated, with the appearance of intensity hot spots in the vicinity of the target. The formation of such hot spots might be interpreted as a possible trade-off between optimization of the intensity and of the intensity gradient at the target position. Comparing the intensity $I$ at the target position when optimizing the CRLB to the intensity $I_{\mathrm{max}}$ obtained after a direct optimization of the intensity on the target, we observe that the intensity ratio $I/I_{\mathrm{max}}$ varies from zero to one [\fig{fig_intensity}(d)]. This confirms that determining the most informative wavefront cannot be reduced to a simple optimization of the intensity at the position of the targeted scatterer. We also show in \fig{fig_intensity}(c) the intensity around the target obtained by minimizing a single CRLB associated with the estimation of both coordinates $\mathcal{C}_{xz} = {\lVert \bm{\mathcal{C}} \rVert}_2$. In that case, we observe that optimizing the CRLB is not strictly equivalent to optimizing the intensity at the target position, but the distribution of $I/I_{\mathrm{max}}$ at the target position becomes strongly skewed towards unity [\fig{fig_intensity}(d)].

In the shot-noise limit, the CRLB scales with the reciprocal of the square root of the number of interactions between the probe field and the target~\cite{giovannetti_advances_2011}, which is here given by the power scattered by the dipole. Since this power is proportional to $|d|^2$ (where $d$ is the dipole induced in the target), on average, we can expect the CRLB to scale with $|d|^{-1}$. The induced dipole can be expressed as the sum of an excitation by the external local field, and a contribution resulting from back action by the environment. This can be written as~\cite{bouchet_quantum_2019}
\begin{equation}
d=\alpha_0 \epsilon_0 E_{\mathrm{exc}}(\bm{r}_0) + \alpha_0 k^2 S(\bm{r}_0,\bm{r}_0) d \; ,
\label{equation_dipole}
\end{equation}
where $S=G-G_0$ is the difference between the Green function in the presence of the medium and the free-space Green function, and $E_{\mathrm{exc}}(\bm{r}_0)$ is the excitation field at the target position, generated by scattering of the incident field by the other scatterers. From \eq{equation_dipole}, we can define a dressed polarizability $\tilde{\alpha}= \alpha_0 [1 - \alpha_0 k^2 S(\bm{r}_0,\bm{r}_0)]^{-1}$ such that $d=\tilde{\alpha} \, \epsilon_0 \, E_{\mathrm{exc}}(\bm{r}_0)$. This simple relation is a convenient starting point to investigate the cases of weakly and strongly scattering targets.

\begin{figure}[b]
\centering
\includegraphics[width=8.5cm]{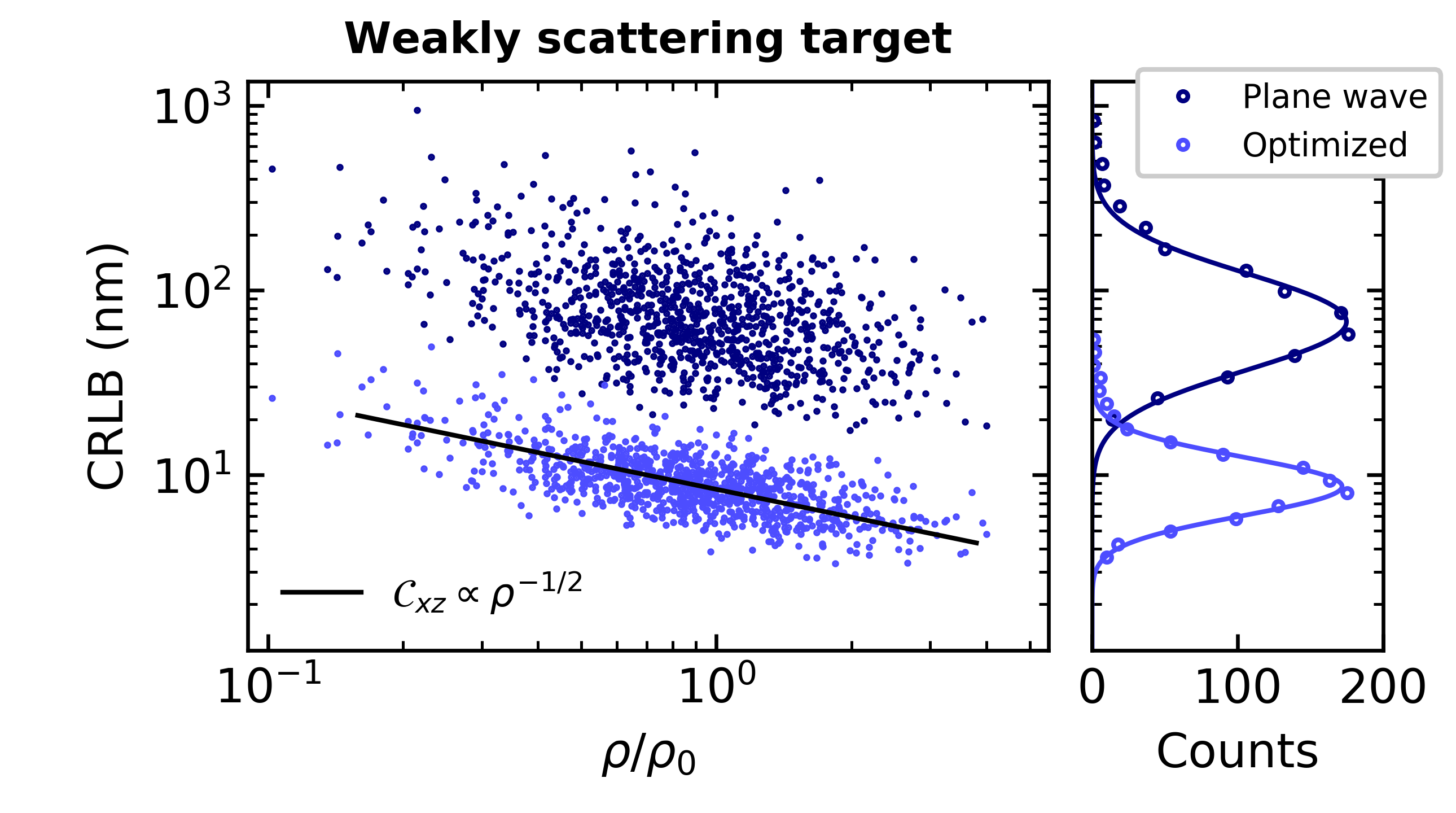}
\caption{Left: $\mathcal{C}_{xz}$ for a weakly scattering target as a function of the normalized LDOS. The black line is a fit to the optimized data by a power law with an exponent of $-1/2$ (a linear regression gives an exponent of $-0.46$). Right: Observed distribution of the CRLB. Log-normal distributions (solid lines) are fitted to numerical observations (data points).}
\label{fig_ldos}
\end{figure}

For a weakly scattering target ($ |\alpha_0 k^2 S(\bm{r}_0,\bm{r}_0)| \ll 1$), back action from the medium is negligible and we can write $\tilde{\alpha} \approx \alpha_0$. In this regime, the strength of the induced dipole $|d|^2$ depends mainly on the intensity of the excitation field at the target position $|E_{\mathrm{exc}}(\bm{r}_0)|^2$. After the optimization of $\mathcal{C}_{xz}$, we observe that the intensity at the target position scales with the LDOS (see Supplementary Section~VI). This is in agreement with a known result related to wave focusing in complex media by time reversal~\cite{carminati_theory_2007}. Consequently, we can expect the CRLB to scale with the reciprocal of the square root of the LDOS at the target position. To prove this assertion, we remove the target and calculate the LDOS $\rho =2k/(\pi c)\im \left[ G (\bm{r}_0,\bm{r}_0) \right]$ at the target position. Introducing the free-space LDOS $\rho_0$, the normalized LDOS at the target position is then expressed by $\rho/\rho_0=1+4 \im \left[S(\bm{r}_0,\bm{r}_0)\right] $. The normalized LDOS can be calculated numerically with the coupled dipole method, using a dipole source located at $\bm{r}_0$. Calculating $\rho/\rho_0$ and $\mathcal{C}_{xz}$ for 1000 configurations, we observe a negative correlation between the LDOS at the target position and the optimized CRLB (\fig{fig_ldos}), characterized by a correlation coefficient of $-0.69$ calculated on log-transformed variables. This result clearly demonstrates that the localization precision of a weak scatterer improves with the value of the LDOS at its position. Furthermore, fitting a power law to numerical observations shows that the optimized CRLB scales with $\rho^{-1/2}$, which is the expected scaling of error in the shot-noise limit. We surmise that an even stronger correlation between $\rho/\rho_0$ and $\mathcal{C}_{xz}$ could be observed with a more complete control of input and output modes.

\begin{figure}[ht]
\centering
\includegraphics[width=8.5cm]{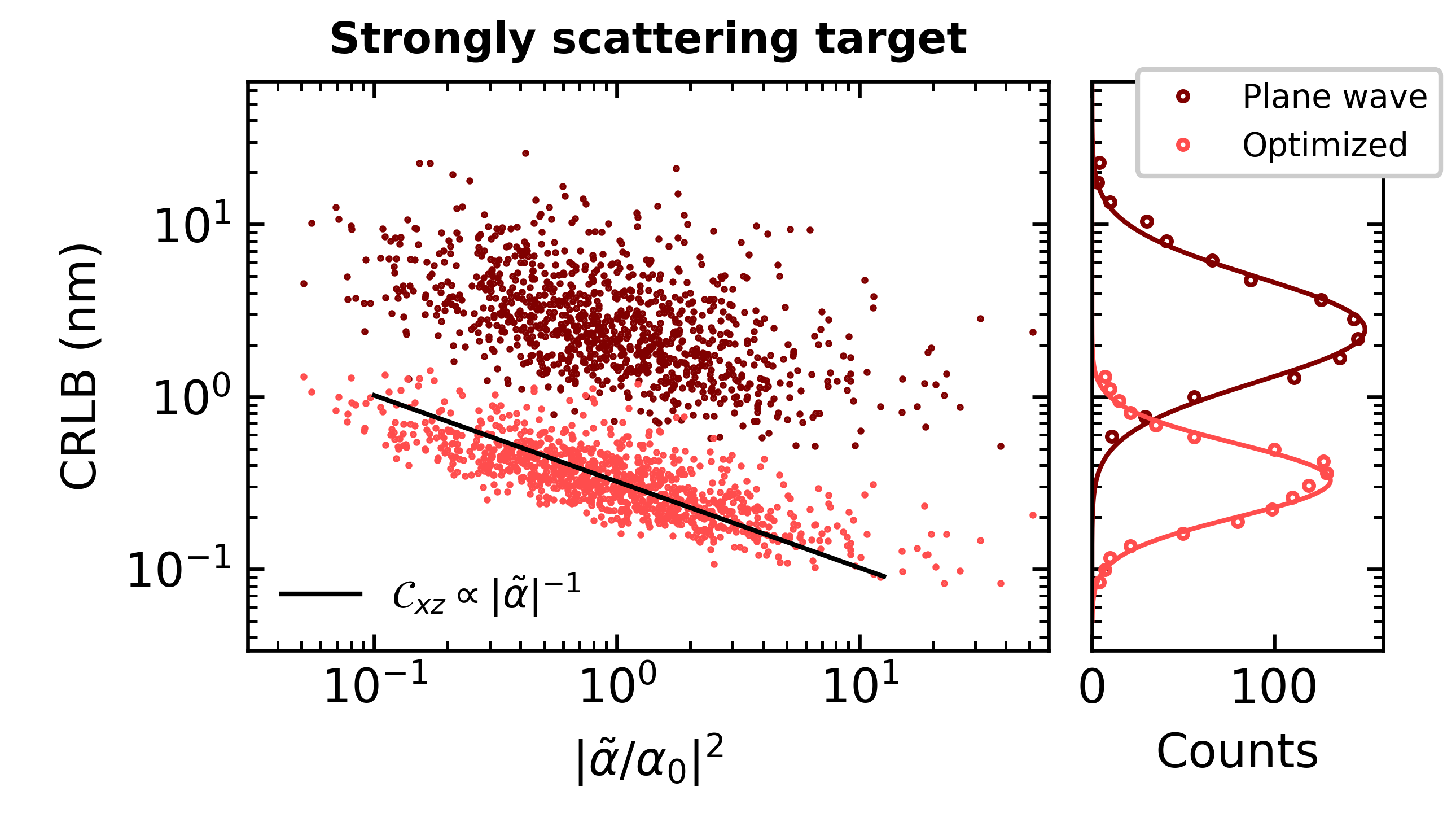}
\caption{Left: $\mathcal{C}_{xz}$ for a strongly scattering target as a function of the normalized effective scattering strength $|\tilde{\alpha}/\alpha_0|^2$. The black line is a fit to the optimized data by a power law with an exponent of $-1/2$ (a linear regression gives an exponent of $-0.36$). Right: Observed distribution of the CRLB. Log-normal distributions (solid lines) are fitted to numerical observations (data points).}
\label{fig_strong_coupling}
\end{figure}

For a strongly scattering target that recurrently scatters the field, the interaction between the target and its environment has to be treated beyond the weak-coupling approximation. To investigate this regime, we set the polarizability of the target on resonance ($\alpha_0=4i/k^2$). The induced dipole depends on the dressed polarizability $\tilde{\alpha}$, which exhibits a pole for $\alpha_0 k^2 S(\bm{r}_0,\bm{r}_0)=1$. Near this pole, we can expect the strength of the induced dipole $|d|^2$ to scale with $|\tilde{\alpha}|^2$ and the CRLB to scale with $|\tilde{\alpha}|^{-1}$. This is confirmed by calculating $|\tilde{\alpha}/\alpha_0|^2$ and $\mathcal{C}_{xz}$ for 1000 configurations, as shown in \fig{fig_strong_coupling}. Indeed, we observe that $|\tilde{\alpha}/\alpha_0|^2$ and $\mathcal{C}_{xz}$ are strongly correlated (with a correlation coefficient of $-0.77$ calculated on log-transformed variables), and that the optimized CRLB roughly scales with $|\tilde{\alpha}|^{-1}$.

In summary, we have introduced a rigorous framework to study and optimize the precision of localization measurements for a subwavelength scatterer in a complex medium. For a weakly scattering target, we have shown that the lower bound on the localization precision depends on the LDOS at the target position. In contrast, the localization precision is driven by the dressed polarizability when recurrent scattering is significant. These results clarify the role of multiple scattering effects for metrology and imaging applications, as well as resulting fundamental limitations~\cite{girard_nanometric_2010}. Our work also opens interesting perspectives in different research areas, for instance to track biological nanoparticles in complex environments~\cite{taylor_interferometric_2019} or to precisely estimate critical parameters of nanomanufactured samples~\cite{alexanderliddle_lithography_2011}. While we have shown that the CRLB can be calculated and optimized using electromagnetic simulations, optimal wavefronts could also be experimentally identified in unknown scattering media by physically modulating the position of the target, for instance using ultrasound-based techniques~\cite{xu_time-reversed_2011,judkewitz_speckle-scale_2013}. Finally, we emphasize that the results are not limited to light waves, and apply to all kinds of waves, for instance to assess and optimize the localization precision of acoustic sources~\cite{kundu_acoustic_2014} or in microwave scattering experiments~\cite{del_hougne_precise_2018}.

\begin{acknowledgments}
The authors thank S. Faez, D. van Oosten and S. Skipetrov for insightful discussions and C. de Kok for IT support. This work was supported by the Netherlands Organization for Scientific Research NWO (Vici 68047618 and Perspective P16-08) and by LABEX WIFI (Laboratory of Excellence ANR-10-LABX-24) within the French Program ``Investments for the Future'' under reference ANR-10-IDEX-0001-02 PSL$^{\ast}$.
\end{acknowledgments}



\onecolumngrid
\clearpage
\beginsupplement
\begin{center}
\textbf{\large Influence of the Local Scattering Environment on the\\Localization Precision of Single Particles\\ \vspace{0.5cm} Supplementary Information}

\bigskip
Dorian Bouchet,$^1$ Rémi Carminati,$^2$ and Allard P. Mosk$^1$\\ \vspace{0.15cm}
\textit{\small $^\mathit{1}$Nanophotonics, Debye Institute for Nanomaterials Science,\\ Utrecht University, P.O. Box 80000, 3508 TA Utrecht, the Netherlands}\\
\textit{\small $^\mathit{2}$Institut Langevin, ESPCI Paris, PSL University, CNRS, 1 rue Jussieu, 75005 Paris, France}
\end{center}
\vspace{1cm}
\twocolumngrid

\section{I. Electrodynamics simulations based on the coupled dipole method}

In this section, we describe the numerical approach used to compute the average value of the intensity in the image plane. The model system is a set of $N_s$ infinite cylinders, confined within an area of transverse dimension $L_z=10 \, \lambda$ and with a larger longitudinal dimension ($L_x=50 \, \lambda$) in order to minimize finite-size effects. A small exclusion radius is defined around the scatterers to prevent them from overlapping. The system is illuminated by an incident field polarized along the longitudinal axis of the cylinders. The scalar wave equation is then solved using a numerical approach based on the coupled dipole method~\cite{lax_multiple_1952_2}, which is an exact formulation in the limit of small cross-sections for the scatterers. Using this model, the field $E(\bm{r}_j)$ at the position of the $j$-th scatterer is expressed by
\begin{equation}
E(\bm{r}_j)=E_0(\bm{r}_j) + k^2 \sum_{\substack{n=0 \\ n \neq j}}^{N_s-1} G_0(\bm{r}_j,\bm{r}_n) \alpha_n E(\bm{r}_n) \; ,
\label{eq_coupled_dipoles}
\end{equation}
where $\bm{r}_n$ is the position of the $n$-th scatterer, $E_0(\bm{r}_n)$ is the incident field at this position and $\alpha_n$ is the polarizability of the scatterer. For 2D systems, the free-space Green function is
\begin{equation}
G_0(\bm{r},\bm{r}') = \frac{i}{4} H_0^{(1)} (k |\bm{r}-\bm{r}'|) \; ,
\end{equation}
where $H_0^{(1)}$ is the Hankel function of the first kind of order 0. \Eq{eq_coupled_dipoles} defines a set of $N_s$ linear equations that are solved using standard computational routines. The field at any position $\bm{r}$ can then be calculated using
\begin{equation}
E(\bm{r})=E_0(\bm{r}) + k^2 \sum_{n=0}^{N_s-1} G_0(\bm{r},\bm{r}_n) \alpha_n E(\bm{r}_n) \; .
\label{eq_coupled_dipoles2}
\end{equation}
Finally, the field in the camera plane is calculated by applying a low-pass filter to the field evaluated at $z=L_z$. Low-pass filtering of the field is performed by convolving it with the product of the cardinal sine function and a Blackman window. In this way, we filter the frequencies higher than $ K_{max}= k \, \mathrm{NA}$ with a transition bandwidth that we set to be on the order of $K_{max}/10$. The numerical aperture of the detection objective is set to $\mathrm{NA}=1$ in the simulations. Assuming that the imaging system has a unitary magnification and choosing a small pixel dimension ($\Delta x=\lambda/10$), the average value for the signal measured by the $i$-th pixel of the camera simply reads $I_i \simeq \Delta x |E_i|^2 $ where $E_i$ is the value of the filtered field at the $i$-th sampling point.

\section{II. Minimum variance unbiased estimator for the linear model}

In this section, we show that we can obtain an explicit expression for an unbiased estimator that reaches the CRLB, in the limit of small parameter variations and for a large number of detected photons. Let us assume that the measured data $\bm{X}$ can be described by a linear model such as
\begin{equation}
\bm{X} = \bm{I} + \bm{J} \bm{d} + \bm{w} \; ,
\end{equation}
where we introduced the intensity vector $\bm{I}=(I_1(\bm{\theta}_0),\dots,I_N(\bm{\theta}_0))^\mathsf{T}$, the displacement vector $\bm{d}=(\Delta \theta_1,\dots,\Delta \theta_K)^\mathsf{T}$, the noise vector $\bm{w}$ and the Jacobian matrix $\bm{J}$ expressed by
\begin{equation}
\bm{J} = \begin{pmatrix}
\partial I_1/\partial \theta_1 & \partial I_1/\partial \theta_2 & \cdots & \partial I_1/\partial \theta_K \\
\partial I_2/\partial \theta_1 & \partial I_2/\partial \theta_2 & \cdots & \partial I_2/\partial \theta_K \\
\vdots & \vdots & \ddots &\vdots \\
\partial I_N/\partial \theta_1 & \partial I_N/\partial \theta_2 & \cdots & \partial I_N/\partial \theta_K \\
\end{pmatrix} \; .
\end{equation}
We assume that the intensity vector $\bm{I}$ and the Jacobian matrix $\bm{J}$ are known. In practice, this can be achieved with a calibration step, which consists of measuring the intensity and its derivative at $\bm{\theta}_0$. Moreover, we assume that the noise vector $\bm{w}$ follows a normal distribution $\mathcal{N}(\bm{0},\bm{C})$, where $\bm{C}$ is the covariance matrix. The normal distribution is indeed a good approximation of the Poisson distribution for a large number of detected photons. Only diagonal terms of the covariance matrix are non-zero, as events detected by different pixels are statistically independent. Thus, the covariance matrix is expressed by
\begin{equation}
\bm{C} = \begin{pmatrix}
I_1(\bm{\theta}_0) & 0 & \cdots & 0 \\
0 & I_2(\bm{\theta}_0) & \cdots & 0 \\
\vdots & \vdots & \ddots &\vdots \\
0 & 0 & \cdots & I_N(\bm{\theta}_0) \\
\end{pmatrix} \; .
\end{equation}
Under these assumptions, the minimum variance unbiased estimator for $\bm{d}$ reaches the Cramér-Rao bound~\cite{kay_fundamentals_1993_2}, and is given by
\begin{equation}
\hat{\bm{d}} = \left( \bm{J}^\mathsf{T} \bm{C}^{-1} \bm{J} \right)^{-1} \bm{J}^\mathsf{T} \bm{C}^{-1} \left( \bm{X}- \bm{I} \right) \; .
\end{equation}
This can be written in the following form:
\begin{equation}
\hat{\bm{d}} = \bm{\mathcal{F}}^{-1} (\bm{\theta}_0) \sum_{i=1}^{N} \nabla_{\bm{\theta}} I_i \left[ \frac{X_i - I_i (\bm{\theta}_0) }{I_i (\bm{\theta}_0) } \right] \; ,
\label{eq_estimator}
\end{equation}
where we introduced the Fisher information matrix $ \bm{\mathcal{F}}(\bm{\theta}_0)$ and the differential operator $\nabla_{\bm{\theta}} = (\partial/\partial \theta_1, \dots, \partial/\partial \theta_K)^\mathsf{T}$. Finally, note that, since only maximum likelihood estimators can be unbiased and efficient, then the estimator expressed by \eq{eq_estimator} is necessarily the maximum likelihood estimator. 

\section{III. Influence of errors on the configuration of the medium}

In this section, we study the robustness of estimations regarding random errors on the position of the dipoles constituting the scattering environment. To this end, we generate $1000$ random configurations of the scattering environment, for three different optical thicknesses (controlled by changing the number of scatterers in the medium). For all calculations, we consider a wavelength $\lambda=633$~nm and an average incident intensity $I_0=10^4$~photons per \textmu m. For each configuration, we determine the optimized wavefront for the coordinate $x_0$, and we construct the estimator defined by \eq{eq_estimator}. We then modify the position of all scatterers constituting the scattering environment according to a Gaussian distribution of variance $\sigma_\textsc{g}^2$, compute the transmitted intensity, and add a random Poisson noise to this intensity. We finally use these numerically-generated data to perform estimations of $x_0$ via the previously-constructed estimator.

\begin{figure}[ht]
\centering
\includegraphics[width=7cm]{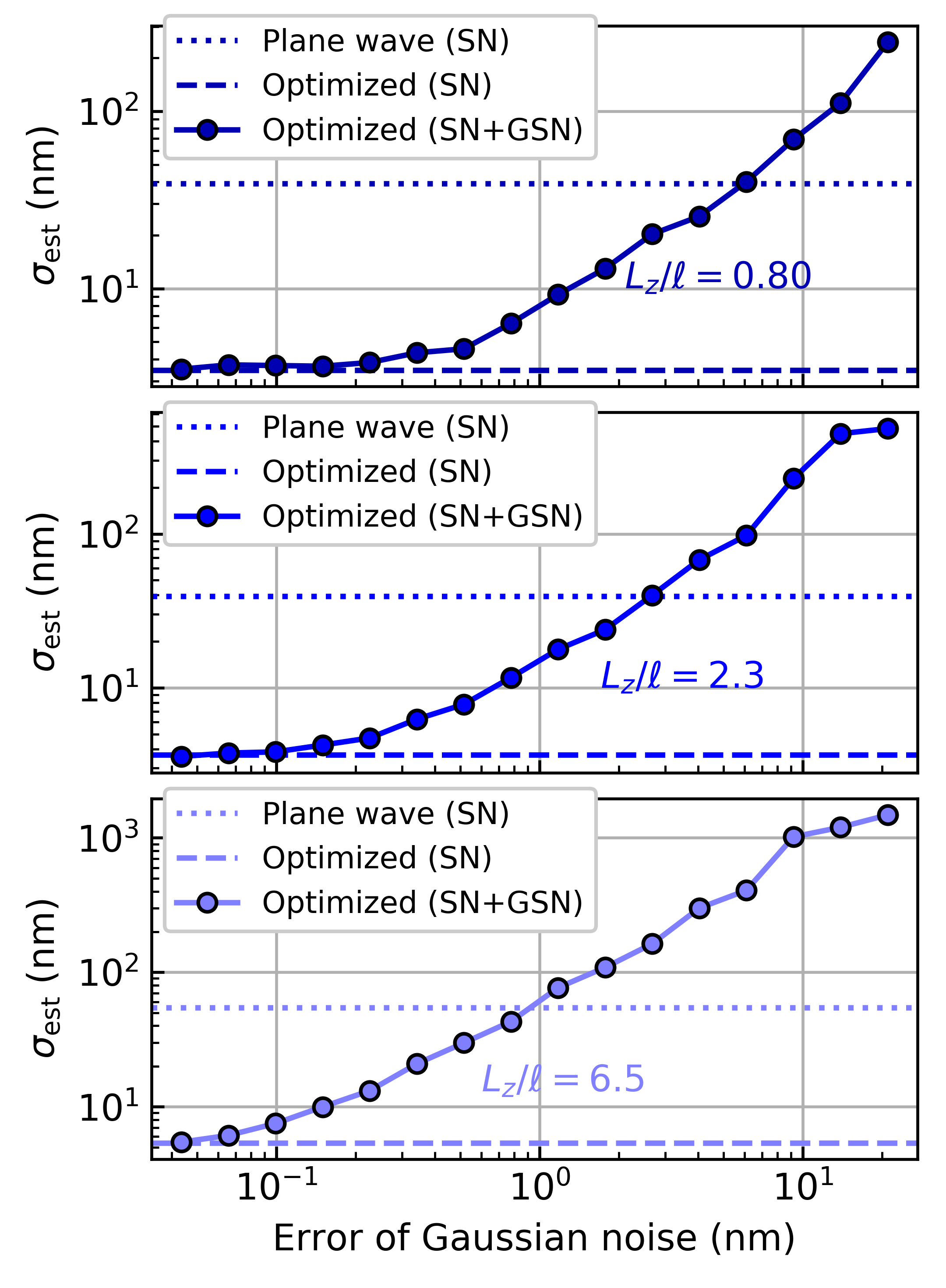}
\caption{Average standard error on the estimates as a function of the standard derivation $\sigma_\textsc{g}$ of the Gaussian noise applied to the dipoles constituting the scattering environment, in the single-scattering regime ($L_z/\ell=0.80$, top panel), in the moderate multiple-scattering regime ($L_z/\ell=2.3$, middle panel) and deeper in the multiple-scattering regime ($L_z/\ell=6.5$, bottom panel). SN: shot noise, GSN: Gaussian structural noise.}
\label{fig_noise_structural}
\end{figure}

We show in \fig{fig_noise_structural} the (arithmetic) average standard error on the estimates $\sigma_\mathrm{est}$ as a function of standard deviation of the structural noise $\sigma_\textsc{g}$ in the single-scattering regime ($L_z/\ell=0.80$, dark blue points), in the moderate multiple-scattering regime ($L_z/\ell=2.3$, medium blue points) as well as deeper in the multiple-scattering regime ($L_z/\ell=6.5$, light blue points). The influence of $\sigma_\textsc{g}$ on the average standard error on the estimates strongly depends on the scattering strength of the environment. Indeed, for the different cases numerically studied, adding a structural noise with a standard error of $1$~nm leads to an average standard error on the estimates of $8.0$~nm (for $L_z/\ell=0.80$), $15$~nm (for $L_z/\ell=2.3$) or $61$~nm (for $L_z/\ell=6.5$).

We can compare these values to the one obtained in the shot-noise limit without structural noise ($\sigma_\textsc{g}=0$). In the case of optimized illumination, the average standard error is $3.5$~nm (for $L_z/\ell=0.80$), $3.7$~nm (for $L_z/\ell=2.3$), and $5.4$~nm (for $L_z/\ell=6.5$). In contrast, plane-wave illuminations leads to an average standard error of $39$~nm (for $L_z/\ell=0.80$), $39$~nm (for $L_z/\ell=2.3$) and $55$~nm (for $L_z/\ell=6.5$). Thus, in the single-scattering regime and in the moderate multiple-scattering regime, optimized illumination with a structural noise of $1$~nm leads to a smaller error than plane-wave illumination with no structural noise. This demonstrate that, with a prior knowledge of the order of $1$~nm as available with current lithography techniques~\cite{alexanderliddle_lithography_2011_2}, studying and optimizing the estimation precision in the shot-noise limit can be directly relevant to the control of manufactured samples, in the single-scattering regime and in the moderate multiple-scattering regime. Only when stronger multiple scattering occurs (for $L_z/\ell=6.5$), it appears that adding a structural noise of $1$~nm degrades the standard error on the estimates as much as using a plane wave instead of an optimized wavefront. Nevertheless, it must be noted that the estimation precision could be further improved by building an estimator that takes into account the incompleteness of the prior knowledge available on the scattering environment, instead of using an estimator based on incorrect prior knowledge.

\bigskip

\section{IV. Log-normal distribution of the Cramér-Rao lower bound}

\begin{figure}[ht]
\centering
\includegraphics[width=6.4cm]{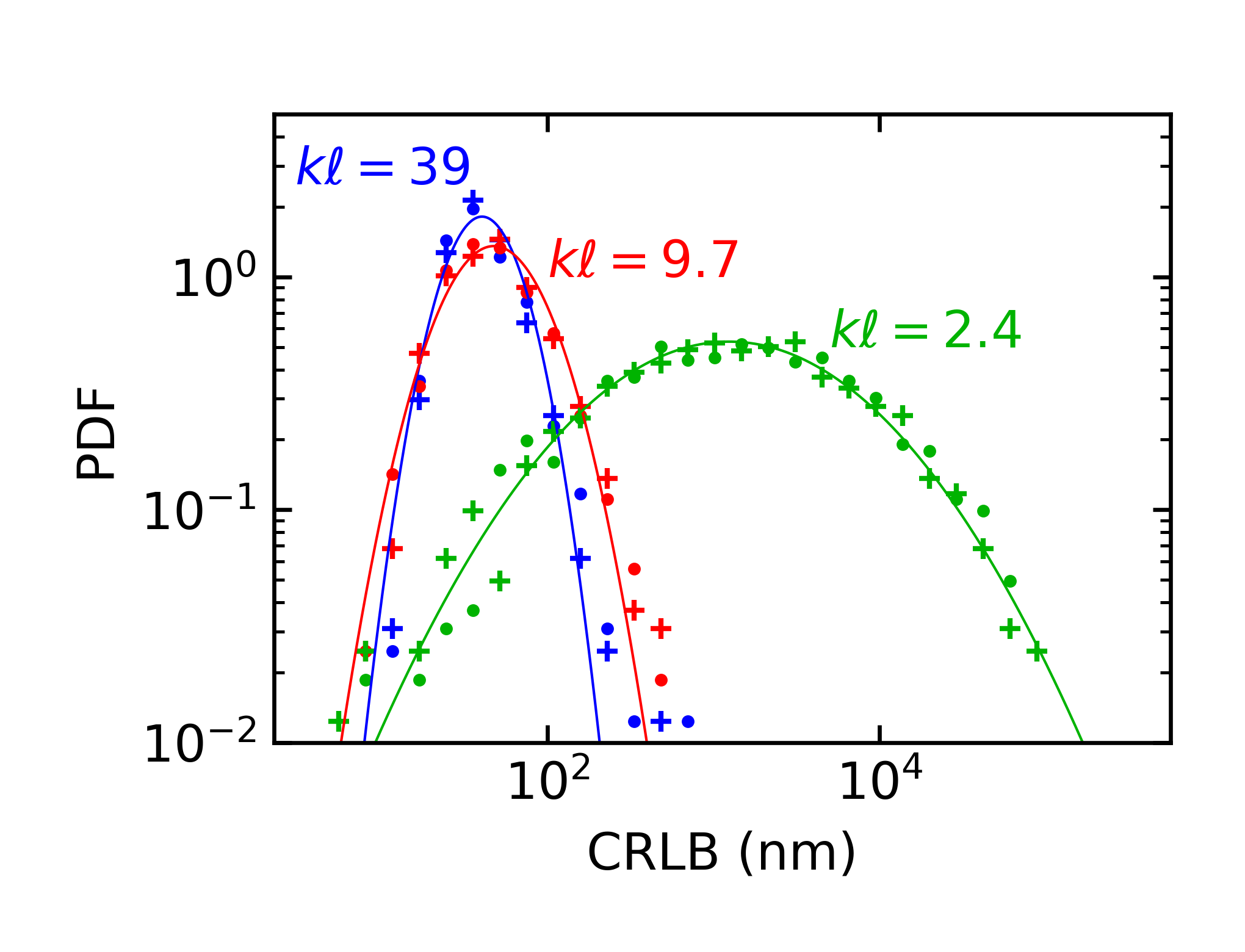}
\caption{Probability density functions followed by the CRLB on the coordinate $x_0$ (data points) and $z_0$ (crosses) for different values of the normalized mean free path. Solid lines are log-normal fits to the data.}
\label{fig_density_function}
\end{figure}

In this section, we show that the CRLB follows a log-normal distribution in the multiple-scattering regime. Indeed, the probability density function followed by the CRLB is correctly fitted by a log-normal distribution for a wide range of scattering mean free path in the multiple scattering regime (\fig{fig_density_function}), thereby justifying to calculate the geometric moments of the distributions rather than the arithmetic ones. As mentioned in the manuscript, decreasing $k \ell$ leads to a broadening of the density function, as well as an increase of the average CRLB. 

\section{V. Convergence of the optimization algorithm}

In this section, we show that the values of the optimized CRLB weakly depend on the initial guess fed to the optimization algorithm, and that the optimized field distributions are strongly correlated. The algorithm that we implemented is based on simulated annealing, which is an adaptation of the Metropolis–Hastings algorithm for approximating the global optimum of a cost function~\cite{kirkpatrick_optimization_1983_2}. The initial guess for the phases of the $N_e$ elements of the SLM is randomly chosen, and the CRLB is iteratively optimized using approximately $700 \times N_e$ function evaluations. Furthermore, at the end of each optimization, we systematically perform a final optimization step using a quasi-Newton method. 

\begin{figure}[ht]
\centering
\includegraphics[width=\linewidth]{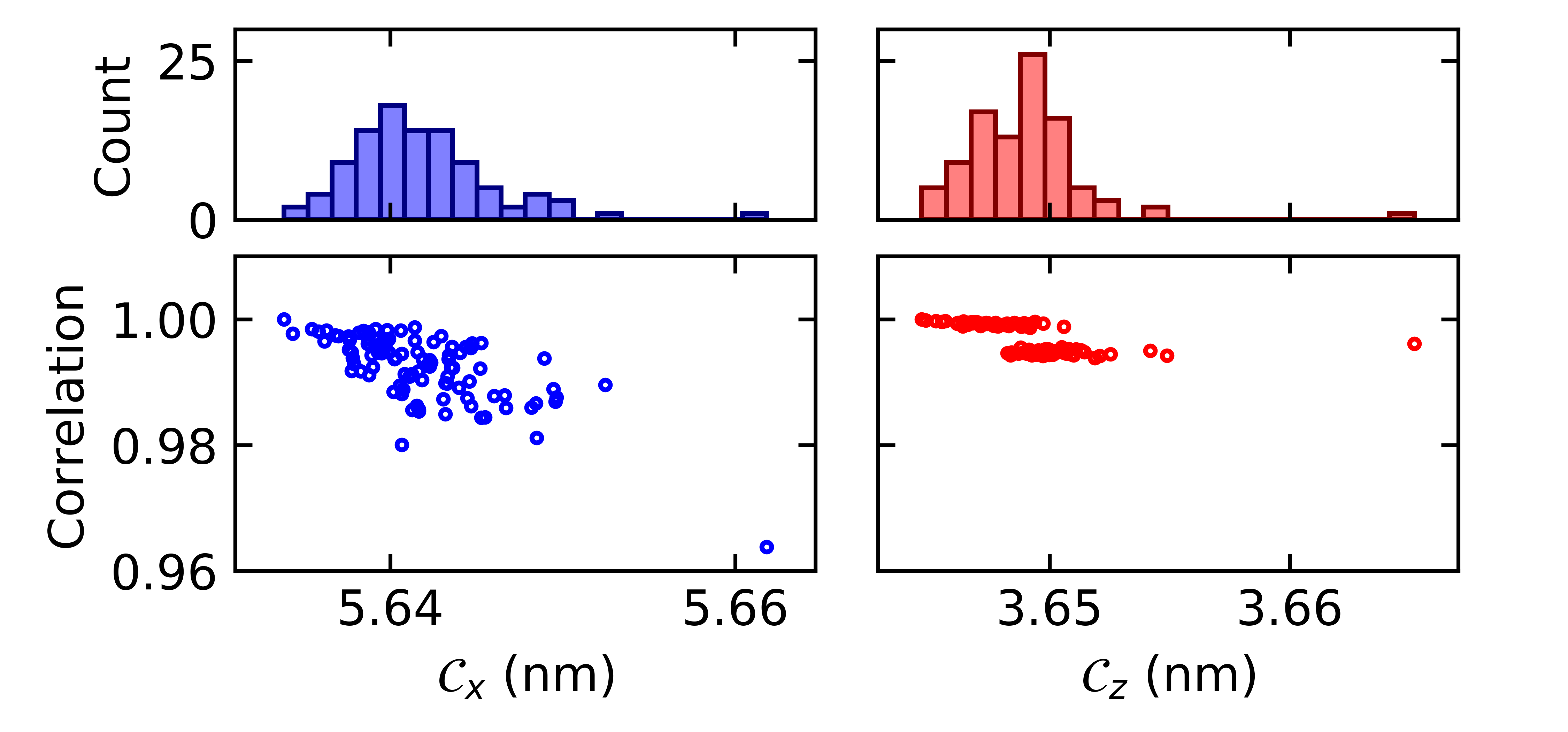}
\caption{Field correlation coefficient as a function of the optimized value of the CRLB for $x_0$ (left panel) and $z_0$ (right panel). The histograms show the distribution of the optimal value found by the optimization algorithm for $100$ different initial guesses. For clarity, three outilers are not shown in the distribution of $\mathcal{C}_z$. The CRLB for these outliers is $3.678$~nm, $3.709$~nm and $3.738$~nm.}
\label{fig_test_sa}
\end{figure}

In order to test the performance of the algorithm, we use the configuration displayed in the manuscript, in the diffusive regime ($k \ell = 9.7$). We successively minimize $\mathcal{C}_{x}$ and $\mathcal{C}_{z}$ using $64$ SLM elements, and we repeat this optimization procedure for $100$ randomly generated initial guesses of the input phases. We can assess the dispersion of the resulting distributions (\fig{fig_test_sa}, upper panels) using the 1-sigma interval defined as $[\mu_g/\sigma_g \, ; \, \mu_g\sigma_g]$ where $\mu_g$ and $\sigma_g$ are respectively the geometric mean and standard deviation of the distribution. The 1-sigma intervals are $[5.638 \, \mathrm{nm} \, ; \, 5.646 \, \mathrm{nm}]$ for the optimization of $\mathcal{C}_x$ and $[3.646 \, \mathrm{nm} \, ; \, 3.651 \, \mathrm{nm}]$ for the optimization of $\mathcal{C}_z$. The dispersion of these distributions is small as compared to the dispersion of the distribution observed when optimizing the CRLB for each coordinate over $1000$ different random configurations, with a 1-sigma interval equal to $[3.412 \, \mathrm{nm} \, ; \, 8.425 \, \mathrm{nm}]$.

In order to determine to what extent the optical modes that are excited are the same for the different solutions, we take the best solution or each coordinate as a reference and we calculate the amplitude of the correlation coefficient for the optimized fields at $z=L_z$ (\fig{fig_test_sa}, bottom panels). The fields associated with the lowest CRLB are highly correlated with the reference field, with a correlation coefficient close to unity. Note that the step-like behavior of the correlation coefficient reflects the possibility for the algorithm to get trapped into a few local optima. Nevertheless, we can see that all the solutions are strongly correlated with the reference field, with a correlation coefficient of at least $0.96$. This indicates that, regardless of the initial guess, the optimization algorithm converges towards similar fields distributions.

\section{VI. Intensity enhancement at the target position}

\begin{figure}[ht]
\centering
\includegraphics[width=8.6cm]{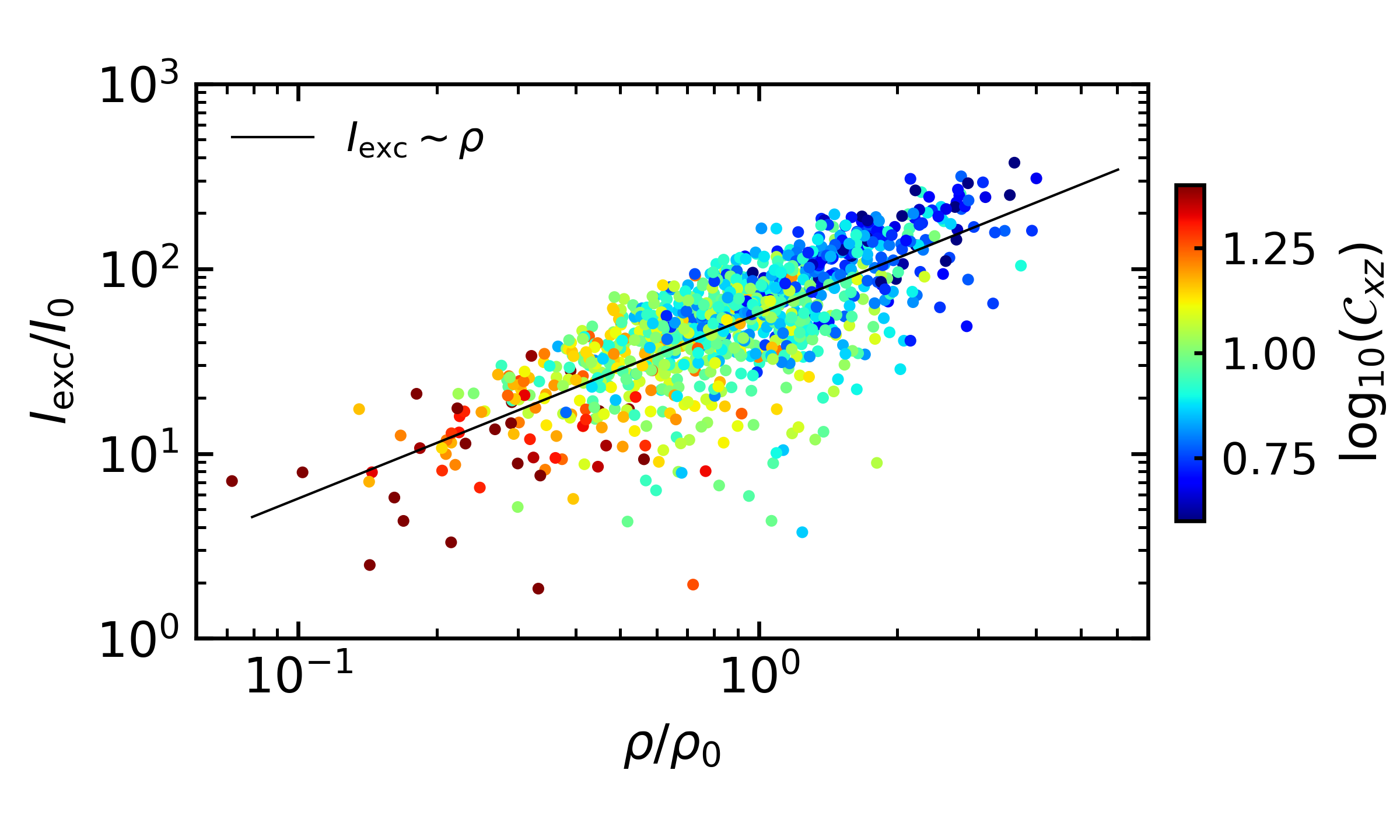}
\caption{Enhancement of the excitation intensity at the position of the target as a function of the normalized LDOS for $k \ell=9.7$ when $\mathcal{C}_{xz}$ is minimized. The color of each point represents the value of $\mathcal{C}_{xz}$, and the black line is a fit to the optimized data by a power law with an exponent equal to $1$.}
\label{fig_scaling_ldos}
\end{figure}

We showed in the manuscript that the CRLB scales with $\rho^{-1/2}$ when $\mathcal{C}_{xz}$ is minimized. This is a consequence of the linear relation between the excitation intensity at the target position and the LDOS. Indeed, the intensity enhancement resulting from the minimization of $\mathcal{C}_{xz}$ scales with the LDOS (\fig{fig_scaling_ldos}), with a correlation coefficient of $0.72$ calculated on log-transformed variables. Furthermore, configurations with a high Cramér-Rao bound are characterized by both a small intensity and a low LDOS. Reversely, configurations with a low CRLB are characterized by both a large intensity and a high LDOS. This confirms that the reduction of the CRLB observed for high LDOS is a consequence of the enhancement of the excitation intensity at the target position.

The observed relation between LDOS and intensity enhancement is in agreement with known results obtained in the context of time-reversal experiments. Such experiments, which provide a method to focus waves into complex scattering media~\cite{mosk_controlling_2012_2}, can be decomposed as two-step processes. During a recording step, a dipole source with constant dipole moment $d_\mathrm{s}$ is located inside a scattering system, and this source emits a field that is recorded on a given surface by a far-field wavefront sensor. During a time-reversal step, a wavefront generator generates on the same surface an incoming field that is the time-reversed replica of the outgoing field previously recorded. It is then known that the power emitted by the source is proportional to the LDOS~\cite{barnes_fluorescence_1998_2,carminati_electromagnetic_2015_2}, and that the intensity of the time-reversed field at the source position is proportional to the product of $|d_\mathrm{s}|^2$ and $\rho^2$~\cite{carminati_theory_2007_2}. In order to relate this latter result to our analysis, which is performed for a constant number of photons injected in the scattering system, we must re-normalized the dipole moment of the source $d_\mathrm{s}$ by the square root of the LDOS, such that the source emits a constant power regardless of its position. In that case, the resulting intensity at the target position scales with the LDOS, in agreement with our numerical results.

The linear relation between $\rho$ and $I_{\mathrm{exc}}$ is strictly valid whenever one has a full control over the input and output modes of the field, which is a necessary condition to achieve a complete time reversal of the field~\cite{carminati_theory_2007_2}. This is not the case in our numerical simulations, which suggests that the correlation between $\rho$ and $I_{\mathrm{exc}}$ could be further improved by performing a more complete control of input and output modes in our simulations.


\end{document}